# Forex Trading Robot Using Fuzzy Logic

Mustafa Shabani [1], Alireza Nasiri [2], and Hassan Nafardi [3]

[1] Master's student in Control Engineering, Department of Electrical and Computer Engineering, Faculty of Engineering, University of Hormozgan, Bandar Abbas, Iran. Also works as a teacher in the Education and Training Organization of Hormozgan Province.

[2] Assistant Professor, Faculty Member, Department of Electrical and Computer Engineering, Faculty of Engineering, University of Hormozgan, Bandar Abbas, Iran.

[3] Head of Technical Support Department, Information Technology Organization of Bandar Abbas Municipality, Iran.

*Abstract--* **In this study, we propose a fuzzy system for conducting short-term transactions in the forex market. The system is designed to enhance common strategies in the forex market using fuzzy logic, thereby improving the accuracy of transactions. Traditionally, technical strategies based on oscillator indicators have relied on predefined ranges for indicators such as Relative Strength Index (RSI), Commodity Channel Indicator (CCI), and Stochastic to determine entry points for trades. However, the use of these classic indicators has yielded suboptimal results due to the changing nature of the market over time. In our proposed approach, instead of employing classical indicators, we introduce a fuzzy Mamdani system for each indicator. The results obtained from these systems are then combined through voting to design a trading robot. Our findings demonstrate a considerable increase in the profitability factor compared to three other methods. Additionally, net profit, gross profit, and maximum capital reduction are calculated and compared across all approaches.**

I. INTRODUCTION

Trading in the forex market presents considerable challenges, where trading volatility is high and daily price fluctuations are significant [1]. Forex traders, due to a lack of understanding of the risks involved in their transactions, often approach the charts with a constant fear of potential losses. Although there are various methods available for entering buy or sell transactions, the matter remains subject to debate, as prices can change randomly and move swiftly [2, 3]. As the forex market operates 24 hours a day, models can be continuously updated and iterated upon, enabling a continuous forecast of forex price trends. This capability holds crucial importance in accurately grasping trading opportunities and mitigating investment risks [4].

In recent years, researchers have put forth various strategies aimed at enhancing the accuracy of trading transactions in order to achieve more precise market predictions, increased profit factors, or improved maximum reduction in transactions. For instance, in [5] a program based on fuzzy logic was introduced to support investment decisions in the forex market. Experimental testing on real forex market data revealed that the utilization of fuzzy logic as an agent's knowledge representation led to more accurate predictions in both upward and downward trends. Similarly, in a [6], proposed robot capable of buying and selling using dynamic price action techniques showcased effective performance compared to the conventional price action technique when trading in digital currency markets. Another study in [7], it involved the comparison of two trading robots, each equipped with three indicators. The first robot incorporated moving average, Bollinger Bands, and Williams indicators, while the second robot utilized Ichimoku, TRIX, and KST indicators. It was found that the first robot outperformed the second robot with a maximum reduction of 17%, as compared to 22%.

When traders are ready to begin their transactions in the forex market, they may find themselves perplexed by the array of trading tools and options available. The forex market offers various types of trades, each presenting different and enticing profit opportunities for traders [8]. A *trading strategy* is a method that simplifies the achievement of a trader's goals. Essentially, it refers to the process by which a trader selects a currency pair and identifies the optimal entry and exit points [6]. In financial markets, there exists a technique called *technical analysis*, which involves predicting the potential behavior of a chart based on past data, such as price movements, trading volume, and more. Technical analysts utilize price charts, volume data, and a range of tools, indicators, and patterns to forecast the probable behavior of a stock [9]. Indicators extract historical information of a trading symbol and present it in a manner that allows traders to assess their own analyses. One advantage of indicators over human interpretation is that human minds are prone to a greater margin of error in mathematical calculations compared to a data-driven system [10].

Traders in the forex market face psychological pressures and high levels of stress when engaging in transactions. Humans are unable to consistently control their emotions and feelings, which can result in reduced focus when calculating the entry points for trades. Consequently, the use of trading robots or computer programs designed based on a set of signals has become increasingly prevalent. These robots can determine when to buy or sell a currency or stock at specific predefined points. The primary objective behind developing these robots is to eliminate detrimental psychological factors from trading. These automated systems are significantly faster and more precise compared to human traders [11].

The RSI indicator was introduced by Wells in 1978 [1]. RSI is a popular technical analysis indicator used to confirm and identify price levels and market reversals. It oscillates between two levels, 0 to 100 [12]. The RSI indicator typically utilizes a 14-day setting for calculations. To compute the RSI value within a 14-day period, the following formula is commonly used:

$$RSI = 100 - \frac{100}{1+RS} \quad (1)$$

$$RS = \frac{Average\ profit}{Average\ loss} \quad .... \quad (2)$$

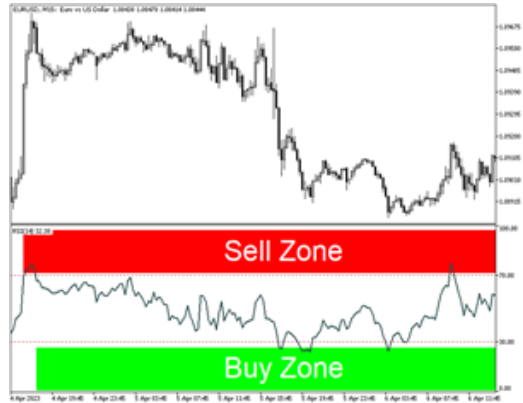

Fig. 1: Buy and Sell Zones

When the RSI indicator surpasses 70 or falls below 30, it indicates strong and significant signs of a market reversal. If the RSI value exceeds 70, it suggests an overbought condition. Conversely, when the RSI drops below 30, it indicates an oversold condition in the market [10, 13] as it can be seen in Fig 1.

The CCI indicator, also known as the "Commodity Channel Index," was introduced by Donald Lambert [1]. This indicator focuses on the strength of recent price trends. Traders use it to make decisions based on the acceleration of buying and selling trends, taking into account the increasing speed of price changes. The CCI fluctuates between -100 and +100. Commonly used time periods for calculating the CCI indicator include 10 days, 20 days, or 30 days [10, 14].

The Stochastic indicator, developed by Dr. Lane in the late 1950s, is renowned as one of the most well-known indicators of price momentum in technical analysis. It oscillates between 0 and 100. The area below the 20 level is considered the oversold or selling-exhaustion zone, while the area above the 80 level is regarded as the overbought or buying-exhaustion zone [1, 15].

Assessing the effectiveness of machine learning strategies leads researchers, scientists, and software developers to often employ various mathematical and economical methods in order to create unpredictable research approaches.

In the realm of financial products, it is crucial to utilize traditional mathematical and economic models for predicting and analyzing sequential data. These models need to be transformed into direct models from non-linear ones. One modeling approach involves the utilization of a fuzzy inference system. There exist multiple methods for constructing a fuzzy inference system, with the most popular ones being proposed by Mamdani and Assilian [16] and Takagi and Sugeno [17]. These systems rely on a series of fuzzy sets to depict the relationship between input variables and their degrees of membership in a set. These sets typically represent attributes that describe the inputs. By considering the elements of the forex market as inputs for the fuzzy system, it becomes possible to delve into a thorough examination of market details [18].

In this study, instead of employing outdated strategies based on classical technical indicators, we aim to design a trading robot using fuzzy logic based on technical indicators, which is expected to yield better performance in forex market transactions. The primary research questions addressed are as follows:

1. Does the utilization of fuzzy logic enhance the profitability factor compared to conventional strategies in the forex market?

2. Can the maximum capital reduction, through the use of fuzzy logic in transactions, be limited to less than 25%?

Moving forward, we will first delve into an explanation of the proposed method in the second section. Subsequently, the third section will present the results of experiments conducted on the proposed method. Finally, the fourth section will provide a conclusion summarizing the findings.

## II. RESEARCH METHODOLOGY

Fuzzy logic was initially introduced by Professor Lotfizadeh in 1965. Fuzzy logic is a reasoning method that exhibits similarities to human reasoning. The objective of a fuzzy system is decision-making. Essentially, fuzzy logic entails the development of an intelligent system capable of interpreting various forms of logical complexity [19]. Fuzzy logic consists of four main components: basic rules, fuzzification, inference engine, and de-fuzzification.

When modeling with Mamdani fuzzy systems, two primary objectives are considered. Firstly, the modeling aims to define, acquire, or reproduce specific input-output functions. Secondly, the modeling seeks to investigate the logical implications of *if-then* rules or sets of *if-then* rules [20].

Mamdani systems possess several appealing characteristics for this purpose, such as high flexibility and universality. They provide a suitable framework for incorporating specialized knowledge in the form of linguistic rules and offer local adaptability to define fuzzy rules within specific regions of the input space [21]. The construction stages of Mamdani fuzzy systems can be outlined as follows in Fig. 2:

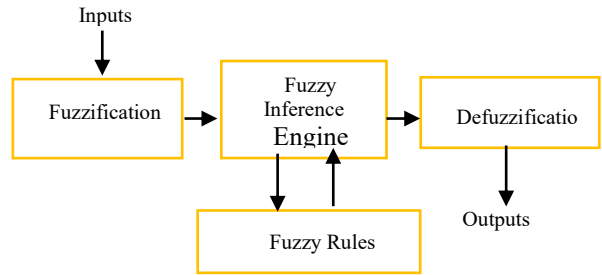

Fig. 2: Mamdani fuzzy system block diagram

The chosen membership functions for the fuzzy model are as follows: where the range (Buy Zone) represents the buying area, the range (Sell Zone) represents the selling area, and the range (Neutral Zone) represents the neutral area as in Fig. 3:

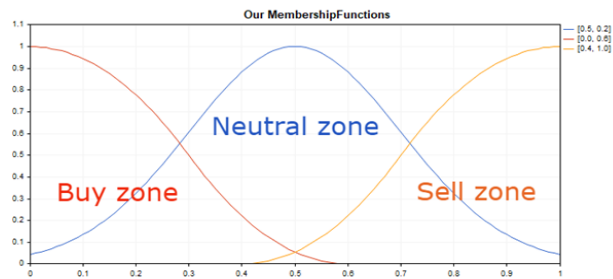

Fig. 3: Selective membership functions

The proposed trading robot comprises four main components: a Mamdani fuzzy system with the RSI indicator, a Mamdani fuzzy system with the CCI indicator, a Mamdani fuzzy system with the Stochastic indicator, and a voting system based on the outputs of the fuzzy systems. Instead of the conventional approach of using indicators and making decisions based on their values, we create a fuzzy system for each indicator. Then, a voting process is conducted among the results obtained from these created systems as in Fig. 4. If the desired conditions are met, the robot will execute a buy or sell transaction; otherwise, it will refrain from entering a trade. The diagram below illustrates the overall structure of the proposed method.

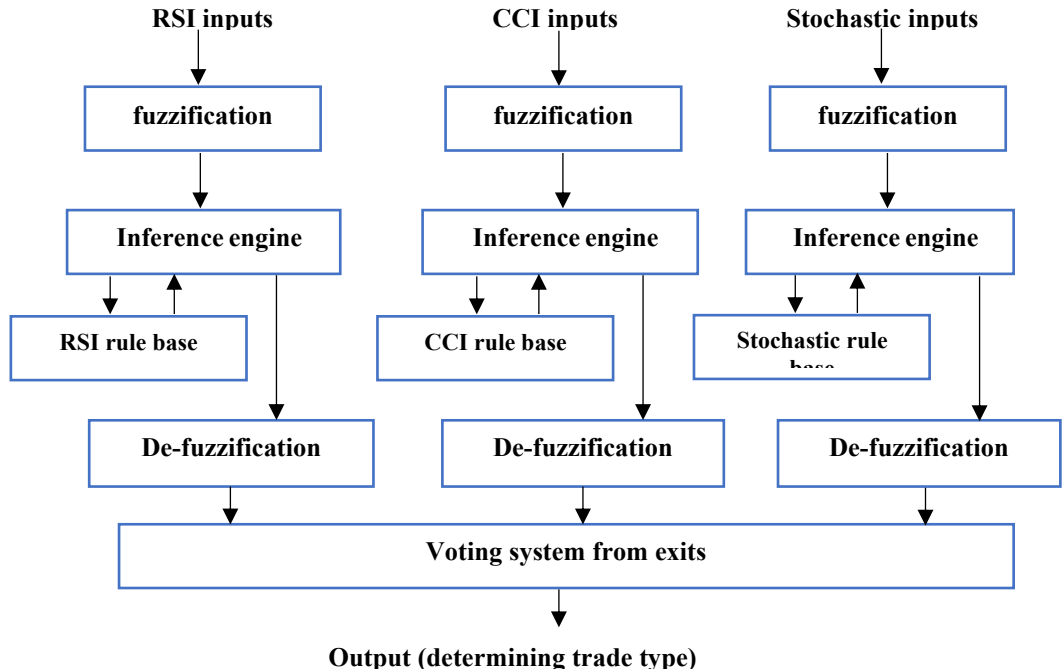

Fig. 4: Output (determining the type of transaction)

Implementation of this system is in the MQL programming environment. For each period used in the RSI and CCI indicators, we consider an input for the fuzzy system. The selected periods for these indicators are 9, 14, and 21 days. Similarly, for each period used in the Stochastic indicator, we consider an input for the fuzzy system. The settings for the Stochastic indicator are as follows:

*First period:  K%: 5, D%: 3, Slow: 3*

*Second period: K%: 14, D%: 7, Slow: 7*

*Third period:K%: 21, D%: 14, Slow: 14*

Then, the fuzzy rules are defined for each system. The fuzzy rules are in the form of *if-then* statements.

**Fuzzy rule base related to the RSI indicator:**

*if (rsi1 is buy) and (rsi2 is buy) and (rsi3 is buy) then (out is buy)*

*if (rsi1 is sell) and (rsi2 is sell) and (rsi3 is sell) then (out is sell)*

*if (rsi1 is neutral) and (rsi2 is neutral) and (rsi3 is neutral) then (out is neutral) ...*

**Fuzzy rule base related to the CCI indicator:**

*if (cci1 is buy) and (cci2 is buy) and (cci3 is buy) then (out is buy)*

*if (cci1 is sell) and (cci2 is sell) and (cci3 is sell) then (out is sell)*

*if (cci1 is neutral) and (cci2 is neutral) and (cci3 is neutral) then (out is neutral) ...*

**Fuzzy rule base related to the Stochastic indicator:**

*if (stoc1 is buy) and (stoc2 is buy) and (stoc3 is buy) then (out is buy)*

*if (stoc1 is sell) and (stoc2 is sell) and (stoc3 is sell) then (out is sell)*

*if (stoc1 is neutral) and (stoc2 is neutral) and (stoc3 is neutral) then (out is neutral) ...*

In total, we have 36 rules for the proposed method, with 12 rules considered for each fuzzy system. Then-after Defuzzification, the following results are obtained:

$$RSI\ Trade\ Signal = \begin{cases} Buy & res < 0.4 \\ Neutral & 0.4 \leq res \leq 0.6 \\ Sell & 0.6 < res \end{cases} \quad (3)$$

$$CCI\ Trade\ Signal = \begin{cases} Buy & res < 0.4 \\ Neutral & 0.4 \leq res \leq 0.6 \\ Sell & 0.6 < res \end{cases} \quad (4)$$

$$STO\ Trade\ Signal = \begin{cases} Buy & res < 0.2 \\ Neutral & 0.2 \leq res \leq 0.8 \\ Sell & 0.8 < res \end{cases} \quad (5)$$

At this stage, the voting system issues the command to enter buying or selling transactions based on the majority vote. For example, if 2 fuzzy systems provide a buy signal and one system provides a neutral signal, the robot will enter a purchase transaction. The same principle applies vice versa, depending on the signals received.

### III. ANALYSIS, EVALUATION, AND TESTING

To analyze the proposed method, we extract a six-month dataset of the EUR/USD currency pair. Using the Metatrader software, we test the proposed fuzzy system and compare it with classical methods that utilize indicators. The obtained results are then compared based on the desired evaluation criteria.

To evaluate trading strategies, various performance metrics assess their effectiveness. One commonly used metric is the *profit factor*. The profit factor represents the ratio between gross profit and gross loss. It serves as an excellent measure to evaluate the performance of a strategy and plays a crucial role in the strategy creation process [45].

$$Invoice\ profit = \frac{gross\ profit}{gross\ loss} \qquad (1)$$

For every trader, it is crucial to know the maximum drawdown of their trading system. You can begin calculating the maximum drawdown as soon as you record the historical peak value of the trading capital. The larger the drawdown (capital loss), the higher the percentage of profit required to recover the capital to its previous level. The maximum drawdown allows traders to manage their trading capital within a logical and reasonable range [46]. The following equation illustrates how to calculate the maximum drawdown accurately:

$$Maximum\ drawdown = \frac{Maximum\ value\ capital - Minimum\ value\ capital}{Maximum\ value\ capital} \qquad (7)$$

To test the developed trading robot, the Metatrader 5 software simulator is utilized. Assuming an initial capital of $10,000, we conduct the experiment during the first six months of 2022, focusing on the EUR/USD currency pair. After testing the proposed method and comparing it with the classical approach of using RSI, CCI, and Stochastic indicators, we have obtained the following results:

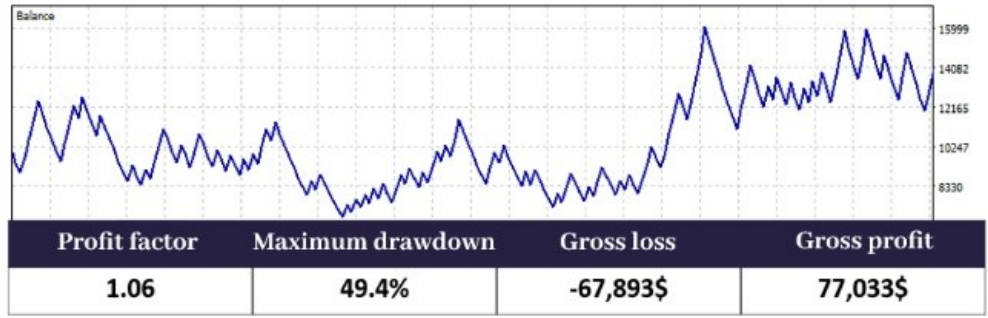

Fig. 5: Test results of the classical RSI system

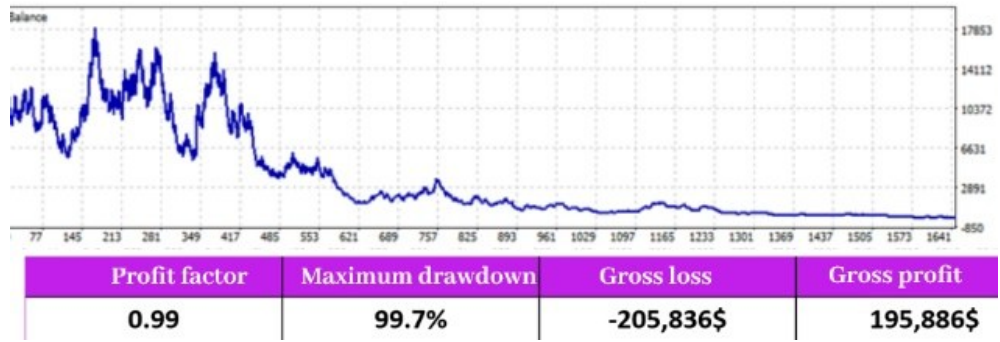

Fig. 6: Test results of the classical CCI system

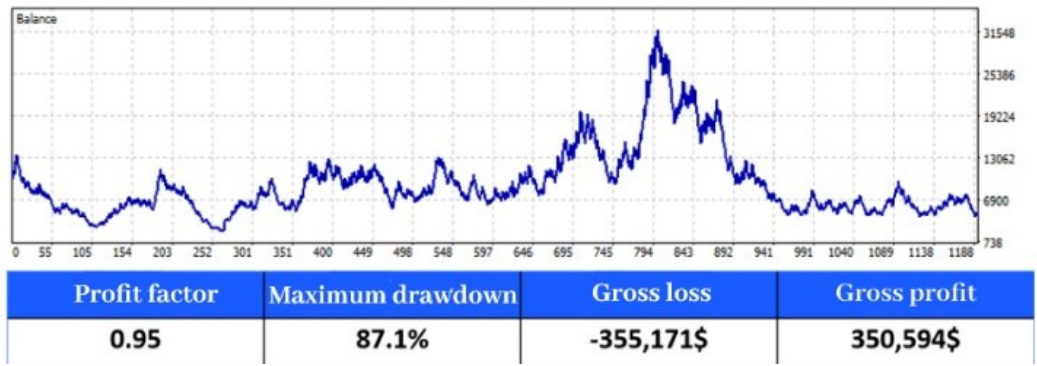

Fig. 7: Test results of the classical Stochastic system

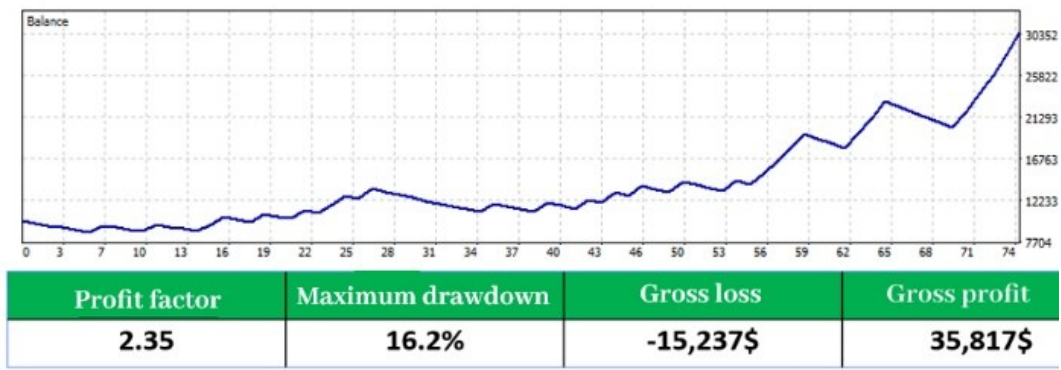

Fig. 8: Test results of the proposed system

As it can be seen in the table 1, the profit factor value is 2.35 for the proposed system, which is higher than to the classic methods (RSI, CCI, and Stochastic indicators). The proposed fuzzy system achieved a profit of 20,579$ in the initial six months of 2022, indicating a higher chance of success.

TABLE I:
PROFIT FACTOR OF THE PROPOSED SYSTEM AND OTHER SYSTEMS

| Proposed System | **2.35** |
|---|---|
| Classical RSI system | **1.06** |
| Classical CCI system | **0.99** |
| Classical Stochastic system | **0.95** |

Also, in the table 2, maximum drawdown value for the proposed fuzzy system and classical indicators have been shown and compared. The best range for maximum drawdown value is less than 25% to have a reasonable trading capital loss. For the proposed method is 16.2% which is considerably less than the others.

TABLE II:
MAXIMUM DRAWDOWN OF THE PROPOSED SYSTEM AND OTHER SYESTEMS

| Proposed System | **16.2%** |
|---|---|
| Classical RSI system | **49.4%** |
| Classical CCI system | **99.7%** |
| Classical Stochastic system | **87.1%** |

## IV. CONCLUSION

In conclusion, this study presented a method to enhance the profit factor in the forex market. Instead of relying on the classical approach using RSI, CCI, and Stochastic indicators, a fuzzy system is developed for each indicator. By employing a voting mechanism based on the obtained results, a trading robot is designed. Simulation results demonstrate that the proposed fuzzy system method significantly outperforms the classical RSI, CCI, and Stochastic indicators in terms of the profit factor. Moreover, it exhibits substantially lower maximum capital reduction during trading compared to other classical methods. The net profit and gross profit are also calculated for all methods. To further advance research in this field, it is recommended to explore optimization algorithms for training the fuzzy system and determining its initial conditions, followed by evaluating its performance.

REFRENCES

[1] Murphy, J.J., Technical analysis of the financial markets : a comprehensive guide to trading methods and applications. First edition ed. 1999, New York: New York Institute of Finance.

[2] Noertjahyana, A., et al. Combination of Candlestick Pattern and Stochastic to Detect Trend Reversal in Forex Market. in 2019 4th Technology Innovation Management and Engineering Science International Conference (TIMES-iCON). 2019.

[3] Malkiel, B.G. and E.F. Fama, EFFICIENT CAPITAL MARKETS: A REVIEW OF THEORY AND EMPIRICAL WORK*. The Journal of Finance, 1970. 25(2): p. 383-417.

[4] Dymova, L., P.Sevastjanov, and K. Kaczmarek, A Forex trading expert system based on a new approach to the rule-base evidential reasoning. Expert Systems with Applications, 2016. p. 1-13.

[5] Korczak, J., M. Hernes, and M. Bac. Fuzzy logic in the multi-agent financial decision support system. in 2015 Federated Conference on Computer Science and Information Systems (FedCSIS). 2015.

[6] Jazayeriy, H. and M. Daryani. SPA Bot: Smart Price-Action Trading Bot for Cryptocurrency Market. in 2021 12th International Conference on Information and Knowledge Technology (IKT). 2021.

[7] Malafeyev, O., et al. Comparative Analysis of Two Trading Robots. in 2022 4th International Conference on Control Systems, Mathematical Modeling, Automation and Energy Efficiency (SUMMA). 2022.

[8] Oztekin, A., K. Best, and D. Delen. Analyzing the Predictability of Exchange Traded Funds Characteristics in the Mutual Fund Market on the Flow of Shares Using a Data Mining Approach. in 2014 47th Hawaii International Conference on System Sciences. 2014.

[9] Sespajayadi, A., Indrabayu, and I. Nurtanio. Technical data analysis for movement prediction of Euro to USD using Genetic Algorithm-Neural Network. in 2015 International Seminar on Intelligent Technology and Its Applications (ISITIA). 2015.

[10] Liu, Z. and D. Xiao, An automated trading system with multi-indicator fusion based on D-S evidence theory in Forex market, in Proceedings of the 6th international conference on Fuzzy systems and knowledge discovery - Volume 3. 2009, IEEE Press: Tianjin, China. p. 239–243.

[11] Zhang, Z. and M. Khushi. GA-MSSR: Genetic Algorithm Maximizing Sharpe and Sterling Ratio Method for RoboTrading. in 2020 International Joint Conference on Neural Networks (IJCNN). 2020.

[12] Rodríguez-González, A., et al. Improving N calculation of the RSI financial indicator using neural networks. in 2010 2nd IEEE International Conference on Information and Financial Engineering. 2010.

[13] Ardimansyah, et al. Moving Average And Relative Strength Index Indicators In Determining Open And Closed Positions On The Metatrader4 Expert Advisor. in 2021 3rd International Conference on Cybernetics and Intelligent System (ICORIS). 2021.

[14] Lambert, D.R., Commodity channel index: Tool for trading cyclic trends. Technical Analysis of Stocks & Commodities, 1983. 1: p. 47.

[15] Izumi, Y., et al. Trading Rules on the Stock Markets using Genetic Network Programming with Candlestick Chart. in 2006 IEEE International Conference on Evolutionary Computation.